\documentclass[fleqn,usenatbib]{mnras}

\usepackage{newtxtext,newtxmath}

\usepackage[T1]{fontenc}

\DeclareRobustCommand{\VAN}[3]{#2}
\let\VANthebibliography\thebibliography
\def\thebibliography{\DeclareRobustCommand{\VAN}[3]{##3}\VANthebibliography}

\usepackage{graphicx}	
\usepackage{amsmath}	
\usepackage{makecell}
\usepackage{threeparttable}

\usepackage{booktabs}

\title[]{Quasi-periodic sub-structure of RRAT J1913+1330}

\author[Z F. Tang et al.]{
Zhenfan Tang$^{1,2}$,
Songbo Zhang$^{1}\thanks{E-mail:sbzhang@pmo.ac.cn}$,
Jieshuang Wang$^{3}\thanks{E-mail:jieshuang.wang@ipp.mpg.de}$,
Xuan Yang$^{1}$,
Xuefeng Wu$^{1,2}\thanks{E-mail:xfwu@pmo.ac.cn}$
\\
$^{1}$Purple Mountain Observatory, Chinese Academy of Sciences, Nanjing 210023, China\\
$^{2}$School of Astronomy and Space Sciences, University of Science and Technology of China, Hefei 230026, China\\
$^{3}$Max Planck Institute for Plasma Physics, Boltzmannstra{\ss}e 2, D-85748 Garching, Germany
}

\date{Accepted XXX. Received YYY; in original form ZZZ}

\pubyear{2024}

\begin{document}
\label{firstpage}
\pagerange{\pageref{firstpage}--\pageref{lastpage}}
\maketitle

\begin{abstract}
Recent findings suggest a universal relationship between the quasi-periodic sub-structures and rotational periods across various types of radio-emitting neutron stars.
In this study, we report the detection of 12 quasi-periodic sub-structures in a rotating radio transient (RRAT) J1913+1330 using the Five-hundred-meter Aperture Spherical Radio Telescope (FAST).
This is the second known RRAT exhibiting quasi-periodic sub-structures.
Our result reinforces the observed relationship between quasi-periodicity and rotational period. 
The polarization analysis reveals that 11 of the 12 pulses exhibit high linear polarization consistent with the quasi-periodic behaviour of the total intensity, while circular polarization with detectable quasi-periodic sub-structures is observed in only three pulses.
No correlation is found between the sub-structure periods and their widths, peak fluxes, or fluences, even under the extremely variable single-pulse energy and morphology observed in J1913+1330.
\end{abstract}



\section{INTRODUCTION}

Soon after the discovery of radio pulsars, sub-millisecond structures known as ``microstructure'' were observed in their single pulses~\citep{1968Natur.218.1122C}.
Analysis of these rapid-intensity variations superimposed on the subpulses identified quasi-periodic structures in second-duration pulsars~\citep{1972ApJ...177L..11H, kramerHighresolutionSinglepulseStudies2002, mitraPOLARIZEDQUASIPERIODICSTRUCTURES2015}. 
Further observation of millisecond pulsars also detected quasi-periodic structures and revealed a robust linear relationship between quasi-periodicity $P_\mu$ and the neutron star rotation period $P_\text{spin}$ (the $P_\mu-P_\text{spin}$ relation)~\citep{deDetectionPolarizedQuasiperiodic2016, liuDetectionQuasiperiodicMicrostructure2022}.   
This correlation indicates a fundamental connection between the mechanism of quasi-periodic sub-structures and pulsar spin.

Recent studies have extended the $P_\mu-P_\text{spin}$ relation by six orders of magnitude in spin period~\citep{kramerQuasiperiodicSubpulseStructure2023}.
Similar quasi-periodic sub-structure phenomena have been discovered in all types of radio-emitting neutron stars, including normal pulsars, millisecond pulsars, RRAT J1918-0449, five out of six radio-emitting magnetars, the newly discovered long-period pulsar J0901-4046, and even the 1091-second pulsating radio source GLEAM-X J162759.5-523504.3~\citep{kramerQuasiperiodicSubpulseStructure2023, caleb2022discovery, hurley2022radio}.
This finding supports the "angular beaming" model as the origin of quasi-periodic sub-structures.
This model suggests the presence of multiple thin flux tubes along the magnetic field lines, where charged particle beams radiate in the direction of propagation, generating an angular radiation pattern~\citep{1979AuJPh..32....9C, gilFrequencyIndependencePulsar1986, petrovaExplanationMicrostructurePulsar2004, 2022ApJ...933..231T, 2022ApJ...933..232T}. 
As the pulsar rotates, the emission structure is sampled over time.
Interestingly, the recent identification of quasi-periodicity in certain Fast Radio Bursts (FRBs) has been proposed to be related to those observed in neutron stars~\citep{2022Natur.607..256C, 2023A&A...678A.149P, 2022RAA....22l4004N, majidBrightFastRadio2021}.

Rotating Radio Transient (RRAT) is a unique subclass of pulsars, typically discovered through single pulse searches~\citep{mclaughlin2006transient}.
To date, over 100 RRATs have been catalogued~\footnote{\url{https://rratalog.github.io/rratalog/} }. 
However, due to the limited number of known RRAT sources and the low pulse rates observed from most of these sources, the classification and theoretical interpretation of RRATs remain challenging~\citep{2006ApJ...645L.149W, burke2012rotating}.
Quasi-periodic sub-structures have been reported in only two RRATs, J1918-0449 and J1913+1330~\citep{chenDiscoveryRotatingRadio2022a, 2024MNRAS.527.4129Z}.
J1913+1330 was among the initial RRAT discoveries~\citep{mclaughlin2006transient}.
In previous observations, only sporadic pulses were detected from RRAT J1913+1330~\citep{mclaughlin2006transient, mclaughlin2009timing, palliyaguru2011radio, 
shapiro2018radio, caleb2019polarization}.
However, recent studies using the Five-hundred-meter Aperture Spherical Radio Telescope (FAST)~\citep{jiang2019commissioning} have found that it exhibits clustered and sequential pulse emissions~\citep{2024ApJ...972...59Z, 2024MNRAS.527.4129Z}.
Unlike some RRATs that exhibit radio emissions similar to ordinary pulsars under more sensitive observations, the energy distribution of J1913+1330 spans three orders of magnitude, resembling the distributions of repeating FRBs~\citep{2021Natur.598..267L, 2022Natur.609..685X}.
The pulse profiles of J1913+1330 are complex and structured, with significant pulse-to-pulse variation.
Exploring the presence of quasi-periodicity in such a special source could help us understand the unusual emission mechanisms of RRATs.

In this study, we conducted a comprehensive analysis and identified 12 quasi-periodic sub-structures in RRAT J1913+1330.
The observation details and data analysis are described in Section~\ref{s2}.
Section~\ref{s3} outlines the properties of the quasi-periodic sub-structures, and Section~\ref{s4} offers a discussion.

\section{OBSERVATIONS AND ANALYSIS}
\label{s2}

\subsection{Observation and Data Reduction}

The data for this study comprises five observations conducted using the FAST between August 17, 2019, and January 14, 2022, totalling 8.9 hours~\citep{2024ApJ...972...59Z}.
Details of these five observing epochs are provided in Table~\ref{t_new}.
The observations employed a 19-beam receiver covering the frequency range of 1000-1500 MHz with 4096 channels~\citep{2019SCPMA..6259502J}.
The sample time was set to 49.153 $\mu\mathrm{s}$ .
On August 17, 2019 (MJD 58712), J1913+1330 was observed with two polarizations, while the remaining four epochs recorded all four polarizations. 
The polarization data were calibrated using the PSRCHIVE software package~\citep{2012AR&T....9..237V}.

The data were processed by two independent search pipelines using the pulsar$/$FRB single pulse searching packages, \emph{\sc presto}\footnote{\url{https://github.com/scottransom/presto}} and \emph{\sc heimdall}\footnote{\url{https://sourceforge.net/projects/heimdall-astro/}}.
Instantaneous and persistent narrow-band radio frequency interference (RFI) were masked out.
The dispersion measure (DM) search trails were set from 165 to 185 $\mathrm{pc~cm^{-3}}$, with a step size of 0.01.
Single pulse candidates with a signal-to-noise ratio (S/N) exceeding 7 were recorded and subjected to visual inspection.
This led to the detection of a total of 1955 pulses within the 5 epochs.
The flux density of each pulse was determined through calibration with noise injected at the beginning of the observations.
The zenith angle-dependent gain of the FAST was also considered~\citep{jiang2020fundamental}.
The fluence was then determined by integrating the flux above the baseline.

\begin{table*}
\caption{Observational information of RRAT J1913+1330 during the five observation epochs. }
\renewcommand\arraystretch{1.2}    
\begin{threeparttable}
\begin{tabular}{lcccccc}
\hline
\hline
Date & Observing Length & Observation used in                 &  number of~\tnote{b}  &  number of  pulses~\tnote{b}               &  number of                \\
 (MJD)    & (hour)           & ~\citet{2024MNRAS.527.4129Z} (hour) &   pulses    &  in~\citet{2024MNRAS.527.4129Z}              &   quasi-periodicities   \\
\hline
  58712 & 0.9 & 0.9  & 141 & 121 & 0 \\ 
  58833 & 3.0 & 2.25 & 685 & 455 & 4 \\ 
  58842 & 1.5 & 1.01 & 395 & 228 & 1 \\ 
  59484 & 1.6 & $-$  & 328 & $-$ & 1 \\ 
  59593 & 1.9 & $-$  & 406 & $-$ & 6 \\ 
\hline
Total      & 8.9 & 4.16 & 1955 & 804 & 12  \\
\hline 
\end{tabular}
  \begin{tablenotes}
        \footnotesize
        \item[a] Our method uses a typical FRB single-pulse search pipeline~\citep{2024ApJ...972...59Z} that is more sensitive to pulses with varying widths.
        \item[b] \citet{2024MNRAS.527.4129Z} employed a typical pulsar single-pulse search pipeline, which involves folding the datasets and searching for peaks in each phase with a constant bin size. As demonstrated by~\citet{2024ApJ...972...59Z}, this approach may not be efficient for RRAT J1913+1330, whose single pulses span a wide range of durations, from 0.15 ms to 17.29 ms—over two orders of magnitude. 
     \end{tablenotes}
\end{threeparttable}
\label{t_new}
\end{table*}

\subsection{Quasi-periodicity Search}

We searched for quasi-periodicity in a total of 1955 single-pulse samples.
For detailed information on single-pulse detections and analysis, refer to~\citet{2024ApJ...972...59Z}.
We applied the auto-correlation function (ACF), a method commonly used for detecting quasi-periodicity for radio pulsars~\citep{taylor1975observations}.
The ACF of a given waveform $I(n)$ at a time lag $k$ is defined as follows:
\begin{equation*}
A(k)=\sum_n I(n+k) I(n).
    \nonumber
\end{equation*}
The ACF of dedispersed time series for each pulse was generated without additional time bins.
Subsequently, we visually inspected the results and identified any ACF exhibiting periodic maxima as candidates.
The time lag of the first peak corresponds to the characteristic $P_\mu$ of the sub-structures.

\subsection{Confidence Evaluation}

To examine the quasi-periodic sub-structure candidates, 
we employed three independent confidence evaluation methods. 
Pulses with a confidence level of $\ge 2\sigma$ in at least two methods were classified as detections with a quasi-periodicity.

\medskip
\medskip
\noindent
\textbf{Rayleigh test: }

The Rayleigh test is a widely used method for evaluating periodicities~\citep{2022Natur.607..256C, kramerQuasiperiodicSubpulseStructure2023}. It employs the parameter: 
\begin{equation*}
Z_1^2=\frac{2}{N}\left[\left(\sum_{i=1}^N \cos \phi_i\right)^2+\left(\sum_{i=1}^N \sin \phi_i\right)^2\right],
\end{equation*}
where $\phi_i$ represents the phase of the time of arrival (ToA) for each sub-structure at a given trial period, and N denotes the number of sub-structures.
To determine the ToA of each sub-structure, we used the \texttt{findpeaks} function from the \textsc{SciPy} module.
For a series of hypothesized periods \( p \), the phases \( \phi \) corresponding to the sub-structures were calculated, and the maximum \( Z \) value was determined. 
For a given pulse, the false alarm probability was calculated by comparing the \( Z \) values obtained from real data with those from randomly generated ToAs. 
The confidence level was then determined accordingly.
The simulated ToAs were created following the method outlined in Equation (9) of~\citet{2022Natur.607..256C}.

\medskip
\medskip
\noindent
\textbf{Statistics $\hat{S}$ test: }

We adopted the method that is used to evaluate the quasi-periodicity of FRBs~\citep{2022Natur.607..256C}.
For a given ToAs, a linear regression was then applied, defined by $t_i = f^{-1} n_i + T_0 + r_i$ to obtain the parameter $\hat{L}[n]$ :
\begin{equation*}
    \hat{L}[n]=\frac{1}{2} \log \left(\frac{\sum_i\left(t_i-\bar{t}_i\right)^2}{r_i^2}\right).
    \nonumber
\end{equation*}
The maximum parameter $\hat{L}[n]$ denoted as the statistic $\hat{S}$:
\begin{equation*}
    \hat{S}=\max _n(\hat{L}[n]).
    \nonumber
\end{equation*}
Significance was determined by comparing the statistic $\hat{S}$ of real data against the distribution derived from randomly generated ToAs.
The ToAs was also generated based on Equation (9) of~\citet{2022Natur.607..256C}.

\medskip
\medskip
\noindent
\textbf{CPS test:}

We also employed method proposed in~\citep{kramerQuasiperiodicSubpulseStructure2023}, which is based on the ACF, called Component Phase Scrambling (CPS).
Initially, the intensity and number of sub-structures in the real data are extracted, followed by the generation of simulated data with random ToA intervals are generated based on this information.
The periodicity of the pulse is then evaluated using the parameter $\mathcal{P}$:
\begin{equation*}
\mathcal{P}=\sum_i \rho\left(P_\mu \cdot i\right), \quad P_\mu \cdot i<l / 2,
\end{equation*}
where the $\rho$ represents the power of local maxima in the ACF and the $l$ denotes the length of the data.
A series of  pulses are simulated base on the number of components, sub-structure width and amplitudes of the real pulse.
The significance is determined by comparing the parameter $\mathcal{P}$ of real data with that of the simulated pulses.

\medskip

We obtained information on the fluence, peak flux, and width of the quasi-periodic sub-structures. 
The flux density and fluence were determined as described in single-pulse search above.
The peak flux is defined as the maximum value of the flux.
The pulse width was determined by dividing the fluence by the peak flux.
The geometric mean, along with the geometric standard deviation factor, was used to determine the range of $P_\mu$.
It is difficult to evaluate the width of sub-structures because many sub-structures in our data have a width of approximately three sample times.
Therefore, we did not provide the value of width of sub-structures.
To investigate correlations between $P_\mu$ and other pulse parameters, we conducted a correlation analysis. 
The Pearson correlation coefficient was utilized to assess linear correlations.

\section{RESULT}
\label{s3}

\begin{figure*}
\centering
\includegraphics[width=1\linewidth]{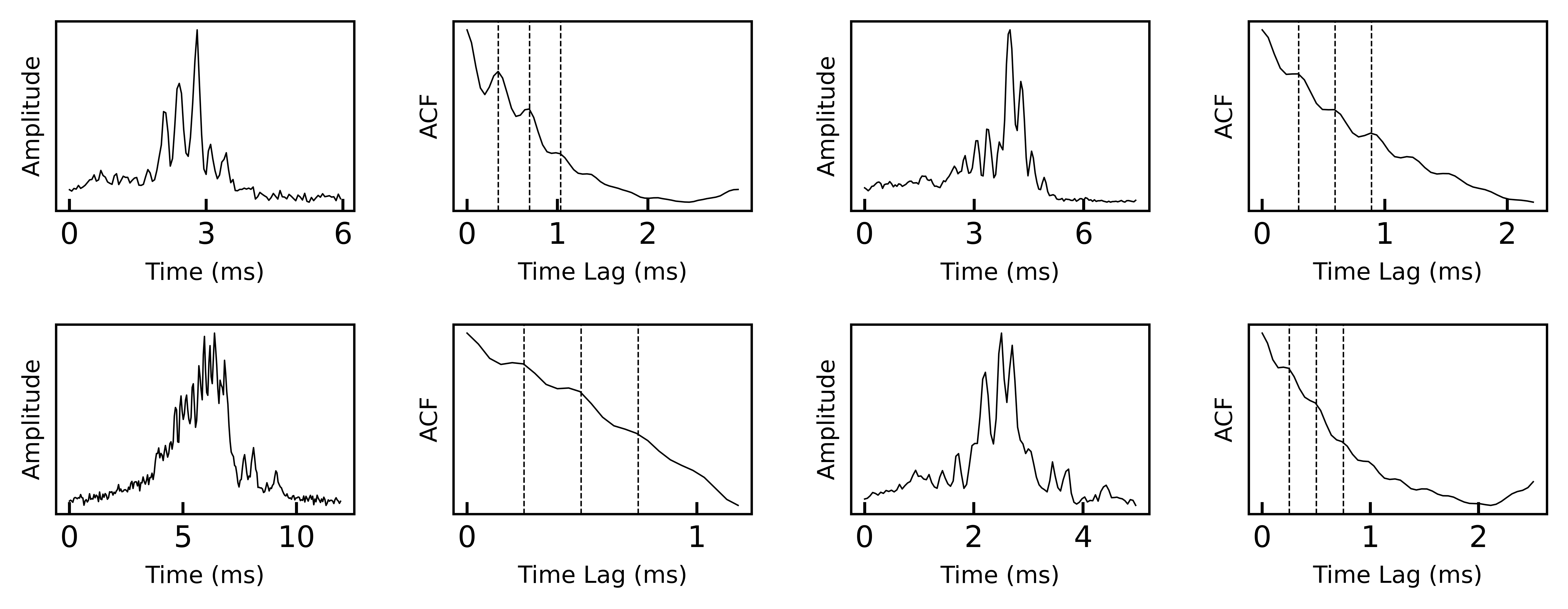}

\caption{
Four examples of quasi-periodic sub-structure of RRAT J1913+1330. 
The sub-panels represent the total intensity as a function of time (left) and the ACF of each pulse (right).
The vertical dashed lines in the ACF panels indicate the detected quasi-periodicities and their harmonics.
}
\label{f1}
\end{figure*}

Thanks to the high sensitivity of FAST, 
more sub-structures of pulses from RRAT J1913+1330 can be detected.
Our analysis of 1955 single pulses led to detection of 12 pulses with quasi-periodic sub-structures.
Figure~\ref{f1} shows examples of these pulses, along with their ACF, which exhibits periodic local maxima.

\begin{figure}
    \includegraphics[width=\columnwidth]{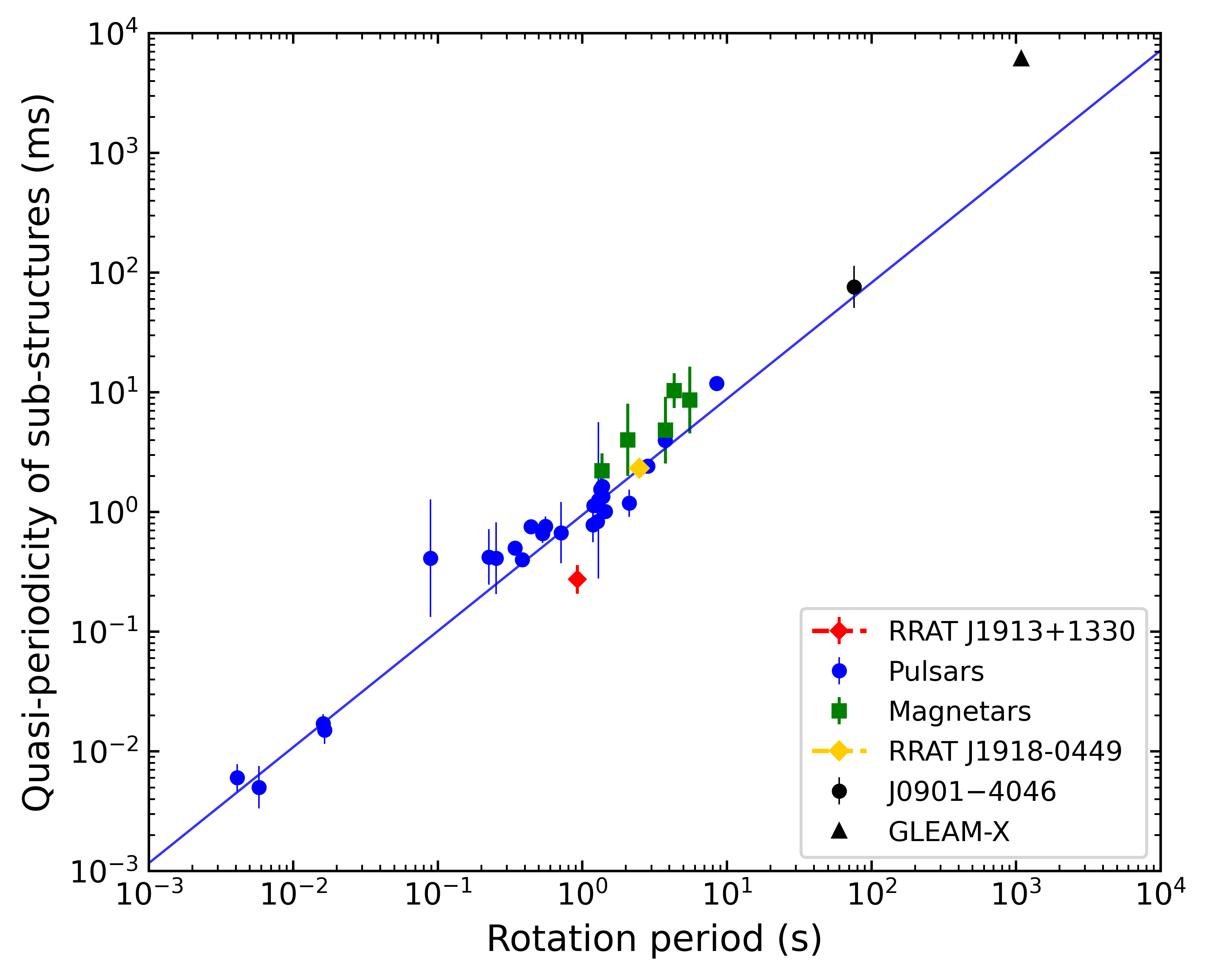}
    \caption{Sub-structure quasi-periodicity of neutron stars against their rotation period, which shows a $P_\mu-P_\text{spin}$ relation. 
    RRAT J1913+1330 is represented by a red diamond.
    Normal and millisecond pulsars are denoted by blue circles.
    Magnetars are illustrated as green squares, whereas the 76-s pulsar, PSR J0901-4046 and pulsating radio source, GLEAM-X J162759.5-523504.3 are marked as black square and triangle.
    Each point on the graph represents the geometric mean, while the error bars are defined by the associated geometric standard deviation.
    The solid line corresponds to the $P_\mu-P_\text{spin}$ relation from recent paper, $P_\mu(\mathrm{ms})=(0.94 \pm 0.04) \times P_\text{spin}(\mathrm{s})^{0.97 \pm 0.05}$~\citep{kramerQuasiperiodicSubpulseStructure2023}. 
    The majority of the values are derived from \protect\cite{kramerQuasiperiodicSubpulseStructure2023}, which obtained from prior studies~\citep{cordes1979coherent, cordesQuasiperiodicMicrostructureRadio1990, lange1998radio, mitraPOLARIZEDQUASIPERIODICSTRUCTURES2015, deDetectionPolarizedQuasiperiodic2016, liuDetectionQuasiperiodicMicrostructure2022}.
    }
    \label{f2}
\end{figure}

\begin{figure}
    \includegraphics[width=0.95\columnwidth]{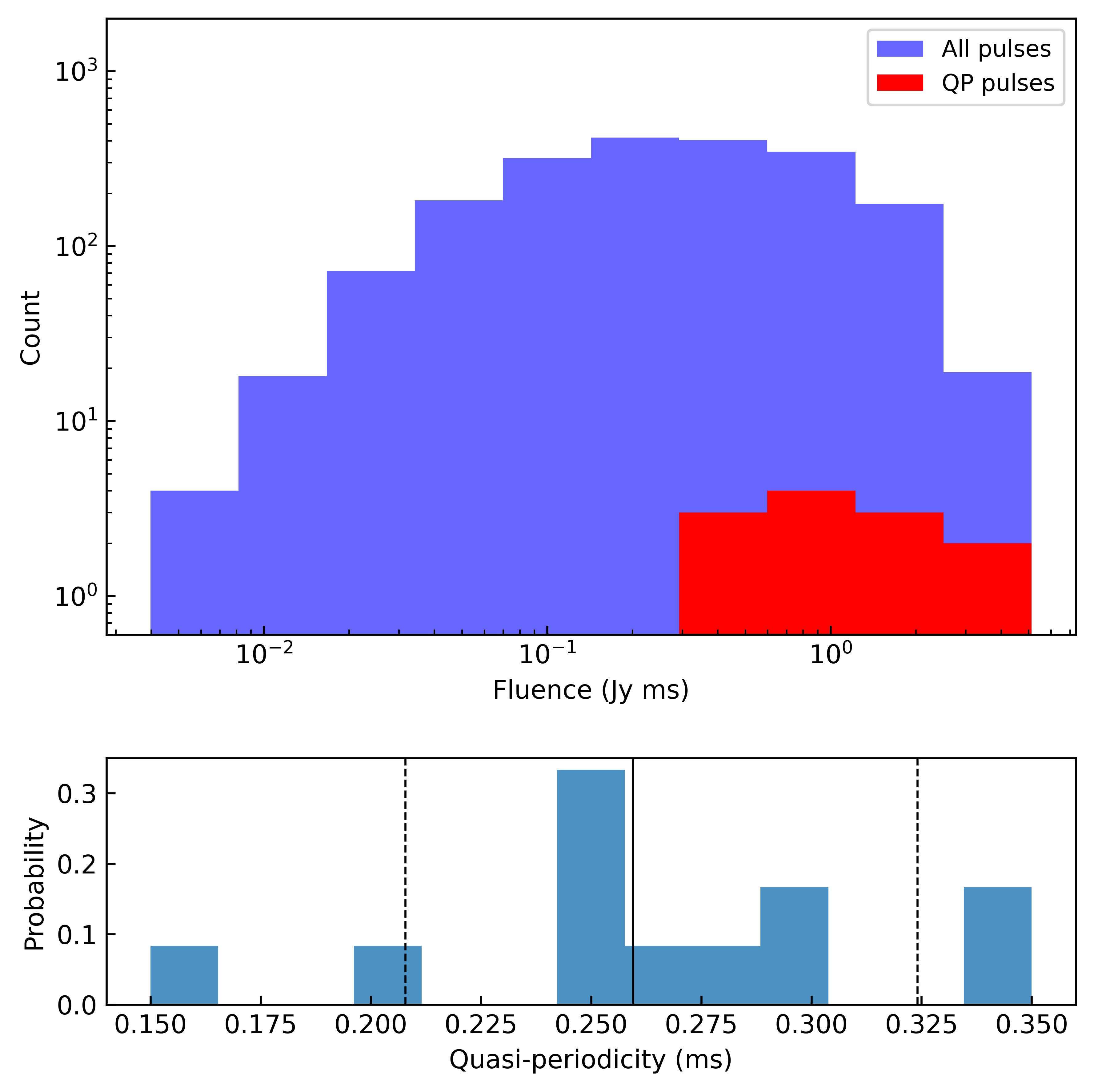}
    \caption{
    The top panel displays the fluence distributions for all pulses (blue) and those exhibiting quasi-periodic sub-structures (red).
    The bottom panel 
    illustrates the distributions of quasi-periodicities $P_\mu$, with a geometric mean of 0.26 ms (solid line) and a geometric standard deviation of 1.25 (dashed lines).
    }

    \label{f3}
\end{figure}

The barycentric ToAs, periods, peak flux, fluences, widths and significance of identified quasi-periodic sub-structures are presented in Table~\ref{t1}. 
The quasi-periodicity is ranging from 0.15 to 0.35 ms.
Pulses with quasi-periodic sub-structures were detected on four out of a total of five epochs of observation.
A comparison of the fluence of quasi-periodic sub-structures with the entire sample of single pulses is depicted in Figure~\ref{f3}.
Quasi-periodic sub-structures appear exclusively in pulses with fluence greater than 0.4 Jy ms.

\begin{figure*}
    \centering
    \includegraphics[width=0.75\linewidth]{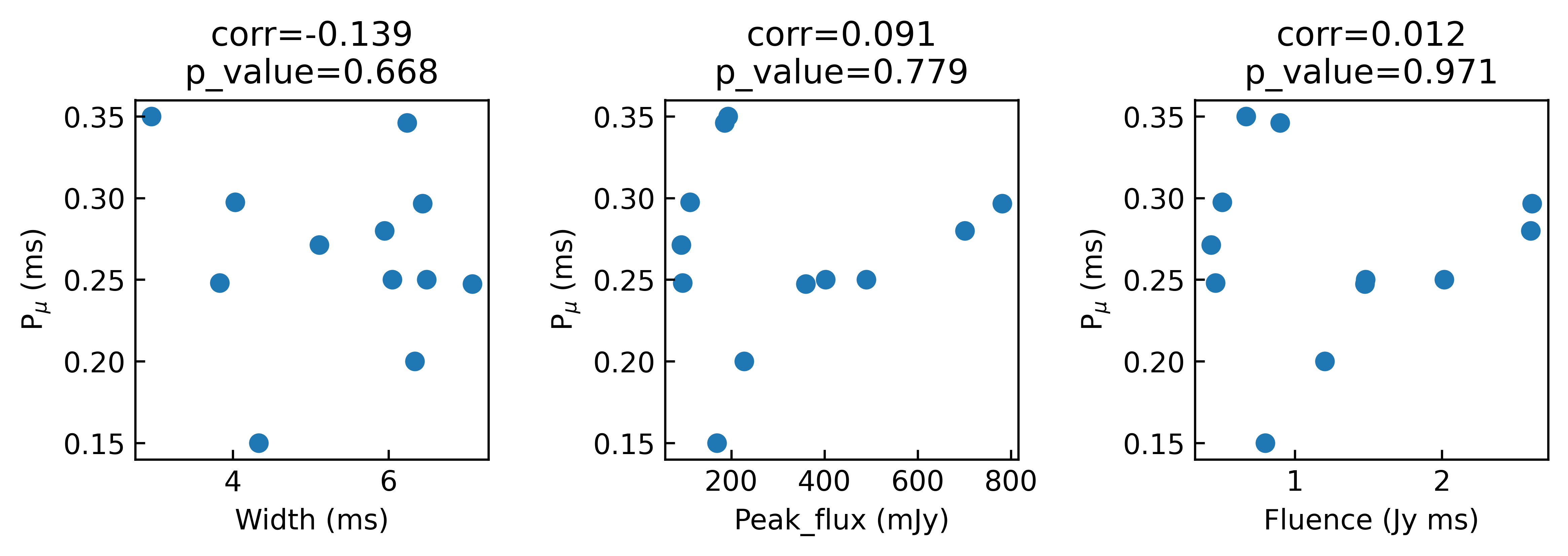}
    \caption{
    Scatter plots of pulse width, peak flux, fluence versus quasi-periodicity.
    The coefficients and p-values of the Spearman correlation are presented in each plot.
    }

    \label{f4}
\end{figure*}

It has been reported that various pulsar parameters, including the period derivative, characteristic age, surface magnetic field, and polarization, do not correlate with $P_\mu$~\citep{mitraPOLARIZEDQUASIPERIODICSTRUCTURES2015}.
RRAT J1913+1330 was recently discovered to have an energy distribution spanning three orders of magnitude, thus we performed correlation tests between $P_\mu$ and the fluences, fluxes and widths. 
The results, illustrated in Figure \ref{f4}, no obvious correlation is found between these parameters and $P_\mu$.

The polarization profiles of all 12 quasi-periodic sub-structures are presented in Figure~\ref{f5}.
Similar to the overall single-pulse samples, the quasi-periodic sub-structures display diverse morphologies and degrees of polarization~\citep{2024ApJ...972...59Z, 2024MNRAS.527.4129Z}. 
The polarization angle generally appears flatter in the quasi-periodic sub-structures. 
We presented the periodic confidence levels of Stokes L and V obtained using the Rayleigh test in Table~\ref{t1}.
High linear polarization is observed in 11 of the 12 pulses, exhibiting the same quasi-periodicity as the total intensity. 
However, circular polarization with detectable quasi-periodic sub-structures is observed in only three pulses.
As shown in Figure~\ref{f6},  in the pulse on the left, both linear and circular polarizations display quasi-periodicity, with their $P_\mu$ consistent with the total intensity.
In contrast, the circular polarization in the right-hand panel of Figure~\ref{f6} completely lacks quasi-periodicity.
For most quasi-periodic sub-structures, the total intensity exhibits higher significance. 
However, the quasi-periodicity of linear polarization for three pulses and circular polarization for two pulses show greater significance than the total intensity. 
This highlights variations in the polarization of quasi-periodic sub-structures between pulses.

\section{CONCLUSION AND DISCUSSION}
\label{s4}

In this paper, we present the detection of quasi-periodic sub-structures from RRAT J1913+1330 observed with the FAST telescope.
Quasi-periodicity has been detected in different types of pulsars, and the $P_\mu-P_\text{spin}$ relation is well-established.
RRAT J1913+1330 has significant pulse variations in profile, fluence, flux, and width~\citep{2024ApJ...972...59Z}, which can provide a unique testing ground for this relation.
In total, we identified 12 quasi-periodic sub-structures in RRAT J1913+1330 with the FAST sample \citep{2024ApJ...972...59Z},
The geometrical mean of the quasi-periodicity $P_\mu$ is 0.26 ms. 
Combining with the rotation period of this RRAT, $\sim$0.923 s, our result is largely consistent with previous findings of the $P_\mu-P_\text{spin}$ relation, as shown in Figure \ref{f2}.

Two types of theoretical models have been proposed to explain the quasi-periodic sub-structures: the ``angular beaming''~\citep{1979AuJPh..32....9C, gilFrequencyIndependencePulsar1986, petrovaExplanationMicrostructurePulsar2004, thompsonRadioEmissionPulsars2022} and ``temporal modulation'' models~\citep{1979AuJPh..32....9C, gilFrequencyIndependencePulsar1986, petrovaExplanationMicrostructurePulsar2004}. 
The $P_\mu-P_\text{spin}$ relation across various types of radio-emitting neutron stars supports the angular beaming model~\citep{deDetectionPolarizedQuasiperiodic2016, kramerQuasiperiodicSubpulseStructure2023}.

We analysed the possible relation between $P_\mu$ and fluence, width, or peak flux, and no statistically significant correlation was found.
This finding suggests that $P_\mu$ is likely independent of these parameters.
In other words, even with the variable single-pulse energy and morphology observed in J1913+1330, the quasi-periodic sub-structures in RRATs remain compliant with the universal $P_\mu-P_\text{spin}$ relation.
This finding appears to support the angular beaming model. 
Since the $P_\mu$ and $P_\text{spin}$ are purely geometrically connected.
However, further observations and theoretical discussions are needed to understand why $P_\mu$ remains independent of fluence, width, or peak flux despite significant variations in single-pulse properties.

The polarization behaviours of 12 quasi-periodic sub-structures of J1913+1330 are similar to those observed in normal pulsars~\citep{mitraPOLARIZEDQUASIPERIODICSTRUCTURES2015}.
Notably, features such as a flat polarization angle and quasi-periodicity in linear polarization closely resemble those found in millisecond pulsars and magnetars~\citep{liuDetectionQuasiperiodicMicrostructure2022, kramerQuasiperiodicSubpulseStructure2023}. 
These shared polarization characteristics strengthen the potential connection in the quasi-periodic emission behaviours across different types of radio-emitting neutron stars.

\begin{figure*}
    \centering
    \includegraphics[width=0.85\linewidth]{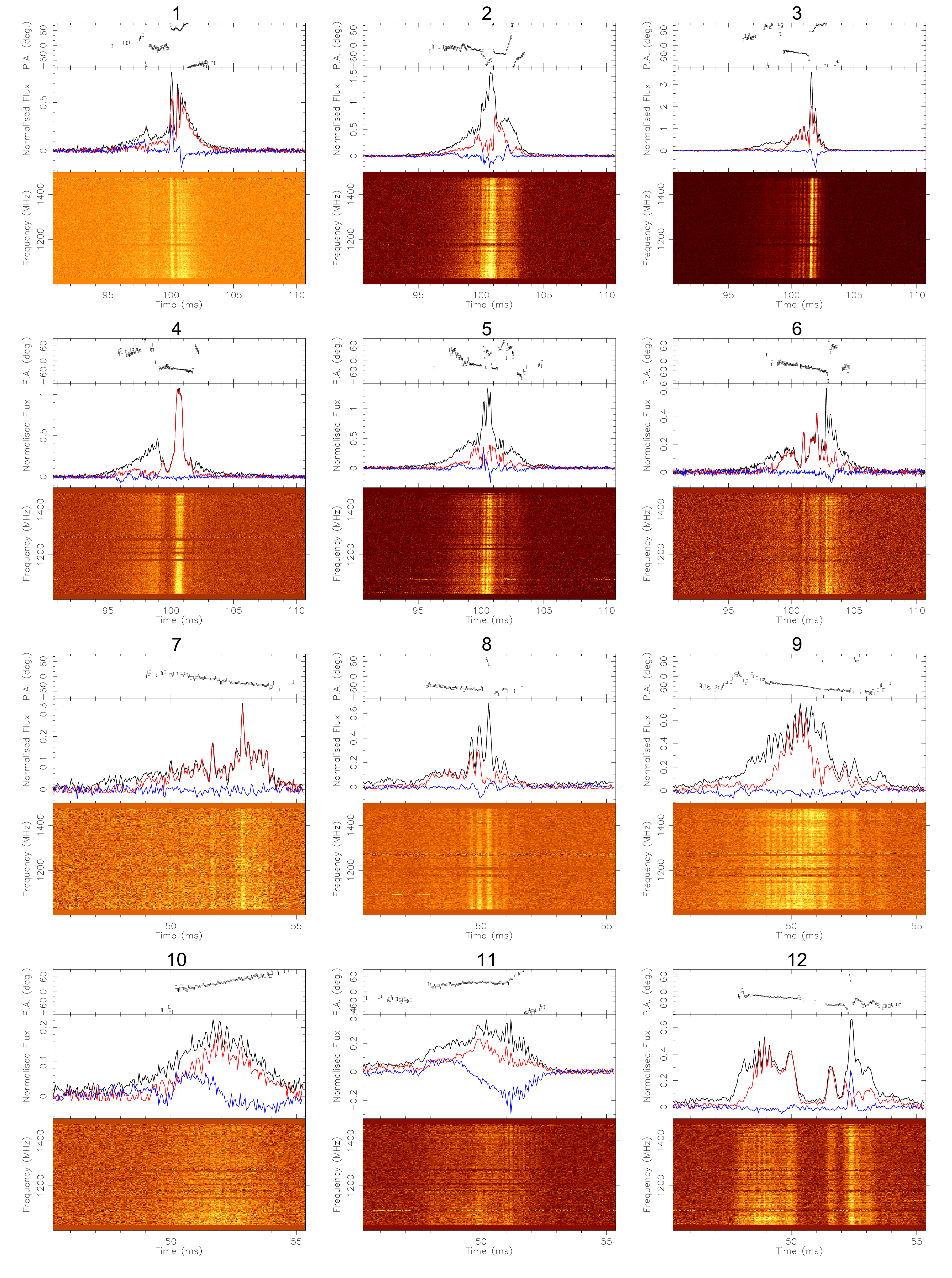}
    \caption{
    Polarization profiles of all 12 quasi-periodic sub-structures of RRAT J1913+1330. 
    In each subplot, the upper panel illustrates the position angle of linear polarization at   a frequency of 1250 MHz.
    The middle panel displays the polarization pulse profile. 
    The black curve represents total intensity, the red curve indicates linear polarization, and the blue curve denotes circular polarization. 
    The lower panel presents the dynamic spectra for the total intensity.
    This is depicted with a frequency resolution of 1.95 MHz per channel and a time resolution of 49.153 $\mu\mathrm{s}$ per bin.
    }
    \label{f5}
\end{figure*}

\begin{figure*}
    \centering
    \includegraphics[width=0.75\linewidth]{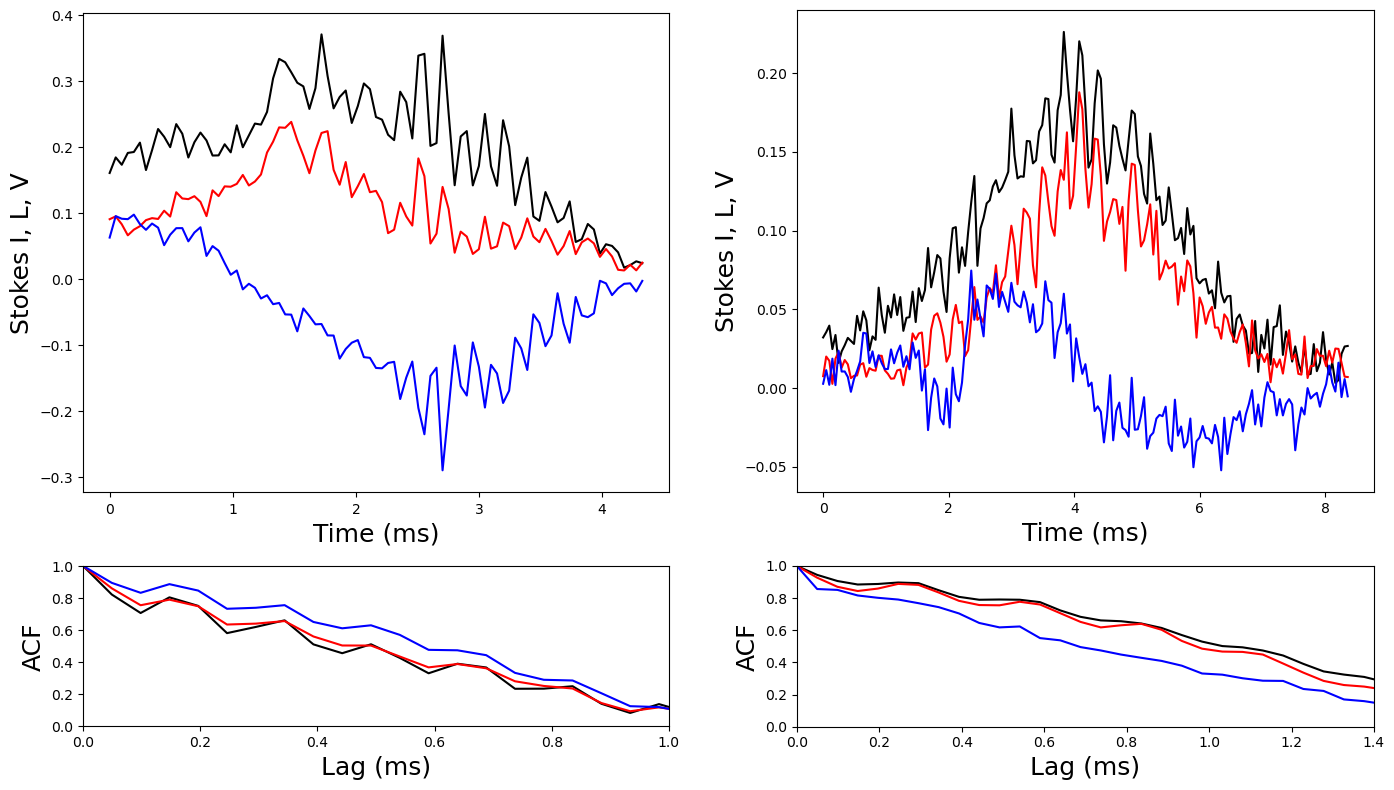}
    \caption{
    Polarization components and their corresponding ACF for a subsample of quasi-periodic sub-structures from RRAT J1913+1330. Black, red, and blue lines represent total intensity, linear polarization, and circular polarization, respectively. In the left figure, the circular polarization component shows significant quasi-periodicity, and the $P_\mu$ of the Stokes \textit{I}, \textit{L}, and \textit{V} are consistent. In the right figure, the circular polarization component lacks quasi-periodicity.}
    \label{f6}
\end{figure*}

As shown in Figure~\ref{f3}, quasi-periodic sub-structures appear exclusively in pulses with fluence greater than 0.4 Jy ms. 
Our examination indicates that this is primarily because pulses below this fluence are heavily contaminated by background noise.
Additionally, we found that this may be attributed to a characteristic of the quasi-periodic sub-structures of RRAT J1913+1330.
These sub-structures are mostly superimposed on the pulse envelopes.  
Some pulsars and magnetars exhibit more discrete quasi-periodic sub-structures, such as certain pulses in~\cite{kramerQuasiperiodicSubpulseStructure2023} Extended Data Fig. 2, that may be detectable at lower fluence levels.  
A small portion of pulses of RRAT J1913+1330 was also found to exhibit discrete sub-structures without the pulse envelopes.
However, they generally did not reach the periodicity confidence threshold we established, except for a few special cases such as Pulse 8 in Figure~\ref{f5}. 
Therefore, for most quasi-periodic sub-structures of RRAT J1913+1330, higher fluence is required to make the relatively weak sub-structures superimposed on the pulse envelopes detectable, resulting in this fluence distribution depicted in Figure~\ref{f3}.  
The underlying causes of such morphological differences in the quasi-periodic sub-structures warrant further investigation.

Notably, as shown in Table~\ref{t_new}, \citet{2024MNRAS.527.4129Z} analysed 4.16 hours of our total 8.9 hours observations and reported 48 quasi-periodicities ranging from 0.306 to 2.818 ms, significantly broader than our 12 detections of around 0.2 ms.
We attribute this discrepancy to differences in data processing and quasi-periodicity analysis methods.
As described by~\citet{2024MNRAS.527.4129Z}, they first folded the datasets with a period of 0.9233913867\,s, and then divided each period into 16,384 bins, yielding a time resolution of 56.53 $\mu\mathrm{s}$. Although their time resolution is only slightly larger than our use of the original sample time of 49.153 $\mu\mathrm{s}$, their data underwent two resampling steps by non-integer factors. 
These steps may have smoothed out fine temporal structures, reducing sensitivity to narrow quasi-periodicities. 
Additionally, ~\citet{2024MNRAS.527.4129Z} employed the Power Spectral Density method to evaluate quasi-periodicity, whereas we used the ACF supplemented by manual verification to ensure that the detected periodicity indeed arose from multiple sub-tructures. 
To further enhance the reliability of our identified sub-structures, we implemented three confidence assessment methods.
With the stricter criteria we applied, which excluded candidates with insufficient confidence, we believe our method more effectively avoids false alarms from major peaks and focuses on narrow quasi-periodic sub-structures. 
Consequently, the final mean quasi-periodicity we obtained is relatively small.

Overall, our results reveal that J1913+1330, as an RRAT with significant pulse-to-pulse variations, exhibits quasi-periodic sub-structures largely consistent with other types of radio-emitting neutron stars in both the $P_\mu-P_\text{spin}$ relation and polarization behaviours.
Next-generation telescopes, such as the SKA and the new proposed FAST array, with a larger field of view or even higher sensitivity, have the potential to identify more and high-confidence quasi-periodic sub-structures from a large sample of RRATs or other pulsating radio source such as white dwarfs.  
This will further test the universality of this relation, and aid in understanding the radiation mechanisms of various types of radio-emitting compact stars.

\begin{table*}
\caption{
Parameters of all identified quasi-periodic sub-structures. 
The ToA is expressed in Modified Julian Date (MJD) that has been corrected to the barycentric time frame.
The confidence levels measured by Statistics $\hat{S}$ test, CPS test and Rayleigh test is denoted as $\sigma_{\hat{S}}$, $\sigma_{CPS}$ and $\sigma_{Z}$.
The last two columns provide the confidence levels of linear and circular polarization using Rayleigh test.
}
\label{t1}
\begin{tabular*}{1\linewidth}{ccccccccccc}
\toprule
No. & MJD & $P_\mu$ ($ms$) & Width (ms) & Peak Flux (mJy)  & Fluence (Jy ms) & $\sigma_{\hat{S}}$ & $\sigma_{CPS}$ & $\sigma_{Z}$ &$\sigma_{Z}$ L & $\sigma_{Z}$ V \\
\midrule
       p1 & 58833.24163501568 & 0.27  & 5.11 & 92.42  & 0.429  & 2.4  & 3.4 & 2.2  & 2.0  & - \\ 
        p2 & 58833.27806988572 & 0.28  & 5.95 & 701.12  & 2.603  & 2.3  & 4.0  & 2.8  & 2.1  & - \\ 
        p3 & 58833.284407778985 & 0.30  & 6.44 & 781.23  & 2.612  & 3.7  & 2.7  & 4.3  & 3.5  & 0.5  \\ 
        p4 & 58833.32603694934 & 0.25  & 6.05 & 402.20  & 1.480  & 2.4  & 2.8  & 2.2  & - & - \\ 
        p5 & 58842.18339870457 & 0.25  & 6.49 & 489.42  & 2.015  & 2.2  & 3.0  & 1.2  & 1.5  & 1.9  \\ 
        p6 & 59484.46675028978 & 0.35  & 6.24 & 186.02  & 0.900  & 2.5  & 2.1  & 2.6  & 2.5  & - \\ 
        p7 & 59593.12646236196 & 0.25  & 3.83 & 95.34  & 0.460  & 3.1  & 2.0  & 2.2  & 2.9  & - \\ 
        p8 & 59593.16148503644 & 0.35  & 2.95 & 192.92  & 0.668  & 2.1  & 2.4  & 1.9  & 2.4  & - \\ 
        p9 & 59593.16186976462 & 0.25  & 7.08 & 359.97  & 1.475  & 2.8  & 3.3 & 2.2  & 0.4  & - \\ 
        p10 & 59593.16307746453 & 0.30  & 4.03 & 111.76  & 0.505  & 3.3  & 3.8 & 3.9  & 2.5  & - \\ 
        p11 & 59593.16510809067 & 0.15  & 4.33 & 169.47  & 0.798  & 2.9  & 3.4  & 3.1  & 2.4  & 3.2  \\ 
        p12 & 59593.16559968235 & 0.20  & 6.34 & 227.54  & 1.203  & 2.0  & 2.3  & 4.8  & 4.5  & - \\
\bottomrule
\end{tabular*}
\end{table*}

\section*{ACKNOWLEDGEMENTS}

This research has been partially funded by the National SKA Program of China (2022SKA0130100), the National Natural Science Foundation of China (grant Nos. 12041306, 12273113,12233002,12003028,12321003), the CAS Project for Young Scientists in Basic Research (Grant No. YSBR-063), the International Partnership Program of Chinese Academy of Sciences for Grand Challenges (114332KYSB20210018), the National Key R\&D Program of China (2021YFA0718500), the ACAMAR Postdoctoral Fellow, China Postdoctoral Science Foundation (grant No. 2020M681758), and the Natural Science Foundation of Jiangsu Province (grant Nos. BK20210998).

\section*{DATA AVAILABILITY}
FAST observational data can be accessed through the FAST data center at \url{http://fast.bao.ac.cn}. Due to the large data volume, the quasi-periodic sub-structures data used in this article shall be shared on reasonable request to the corresponding author.


\bibliographystyle{mnras}
\bibliography{manual}


\bsp	
\label{lastpage}
\end{document}